\documentclass{article}
\usepackage{spconf}
\usepackage{multirow}
\usepackage{booktabs,caption}
\usepackage[flushleft]{threeparttable}
\usepackage[export]{adjustbox}

\title{Scenario Aware Speech Recognition:\\
Advancements for Apollo Fearless Steps \& CHiME-4 Corpora}
\name{Szu-Jui Chen, Wei Xia, John H.L. Hansen \thanks{This project was funded, in part, by NSF-CISE Award 2016725, and partially by the University of Texas at Dallas from the Distinguished University Chair in Telecommunications Engineering held by J. Hansen.}}
\address{Center for Robust Speech Systems (CRSS), University of Texas at Dallas, TX 75080\\ \{szujui.chen,wei.xia,john.hansen\}@utdallas.edu}
\begin{document}
%
\maketitle
\begin{abstract}
In this study, we propose to investigate triplet loss for the purpose of an alternative feature representation for ASR. We consider a general non-semantic speech representation, which is trained with a self-supervised criteria based on triplet loss called TRILL, for acoustic modeling to represent the acoustic characteristics of each audio. This strategy is then applied to the CHiME-4 corpus and CRSS-UTDallas Fearless Steps Corpus, with emphasis on the 100-hour challenge corpus which consists of 5 selected NASA Apollo-11 channels. An analysis of the extracted embeddings provides the foundation needed to characterize training utterances into distinct groups based on acoustic distinguishing properties. Moreover, we also demonstrate that triplet-loss based embedding performs better than i-Vector in acoustic modeling, confirming that the triplet loss is more effective than a speaker feature. With additional techniques such as pronunciation and silence probability modeling, plus multi-style training, we achieve a +5.42\% and +3.18\% relative WER improvement for the development and evaluation sets of the Fearless Steps Corpus. To explore generalization, we further test the same technique on the 1 channel track of CHiME-4 and observe a +11.90\% relative WER improvement for real test data.
\end{abstract}
\vspace{-0.8em}
\section{Introduction}
Significant progresses in automatic speech recognition (ASR) have taken place in recent years. Today, ASR systems are utilized in our daily lives, where a diverse range of recognition scenarios that contain distinct background acoustic conditions are observed. However, modern ASR systems are still struggling to effectively overcome noise levels and adverse background conditions, leading to unsatisfactory recognition results in daily use. It is suggested that this could be caused by less effective ASR acoustic modeling based on Mel-Frequency Cepstral Coefficients (MFCC) or log-mel filterbanks (FBANK) energies. Such features are sensitive to noise \cite{bhattacharjee2016statistical}, and these systems are generally trained for a specific use case and sensitive to test mismatch. Performance can further degrade for distant talking situations, where signal energy is lower, reverberation is possible, and environment signal-to-noise ratio (SNR) is lower. As a result, it is necessary to create a solution to minimize the influences from changing background acoustic conditions.

In the past, methods have been proposed to address the problem of noisy speech recognition \cite{gong1995speech}. Most focus on feature enhancement \cite{yu2008minimum}, or model adaptation \cite{seltzer2010acoustic}. One proposed method is based on factor-aware training. Such a technique introduces factors including noise \cite{seltzer2013investigation}, speaker \cite{tan2016speaker}, and/or room characteristics \cite{giri2015improving} into the training of deep neural networks (DNN) as auxiliary information. This added supporting information serves as a factor-dependent bias to the DNN which causes the output of the DNN to depend on the individual factor values. The most well-known example is the i-Vector that was originally proposed for speaker recognition \cite{dehak2010front}. Here, it is possible for us to apply it as speaker and channel representations in factor aware training.

To address diversity in acoustic characteristics, we propose adding a feature to model the acoustic characteristics, such as channel distortions and environmental noise types, in the audio. The goal here is to make the acoustic model aware of this available information, which can be summarized as a "scenario" that exists in the audio. This idea needs either several good representations for each classifiable factor, or an exceptional representation that can suitably distinguish a specific acoustic context.

Past studies have explored triplet loss as a means for improving speech technology, specially for speaker ID \cite{zhang2018text, zhang2019utd}. However, to the best of our knowledge, triplet loss studies have not been explored in acoustic modeling for ASR. In this study, as motivated by past efforts in speaker recognition \cite{zhang2018text, zhang2019utd}, we employ a triplet-loss based representation generated by TRIpLet Loss network (TRILL) \cite{shor2020towards} for speech recognition. In that network, a subset of the AudioSet \cite{gemmeke2017audio} that possesses the speech label is used for training in a self-supervised manner. Since the AudioSet corpus is a large dataset for general audio machine learning with general audio speech tags, it is useful due to size and scope. As a result, the triplet-loss based representation is expected to learn generalization for audio. The technique developed in \cite{jansen2018unsupervised} was used to allow the network the ability to represent segments in audio that are closer in time to be closer in the embedding space. Details are presented in Sec.\ref{sec:TRILL}.

\begin{figure}[h]
  \centering
  \includegraphics[width=\linewidth]{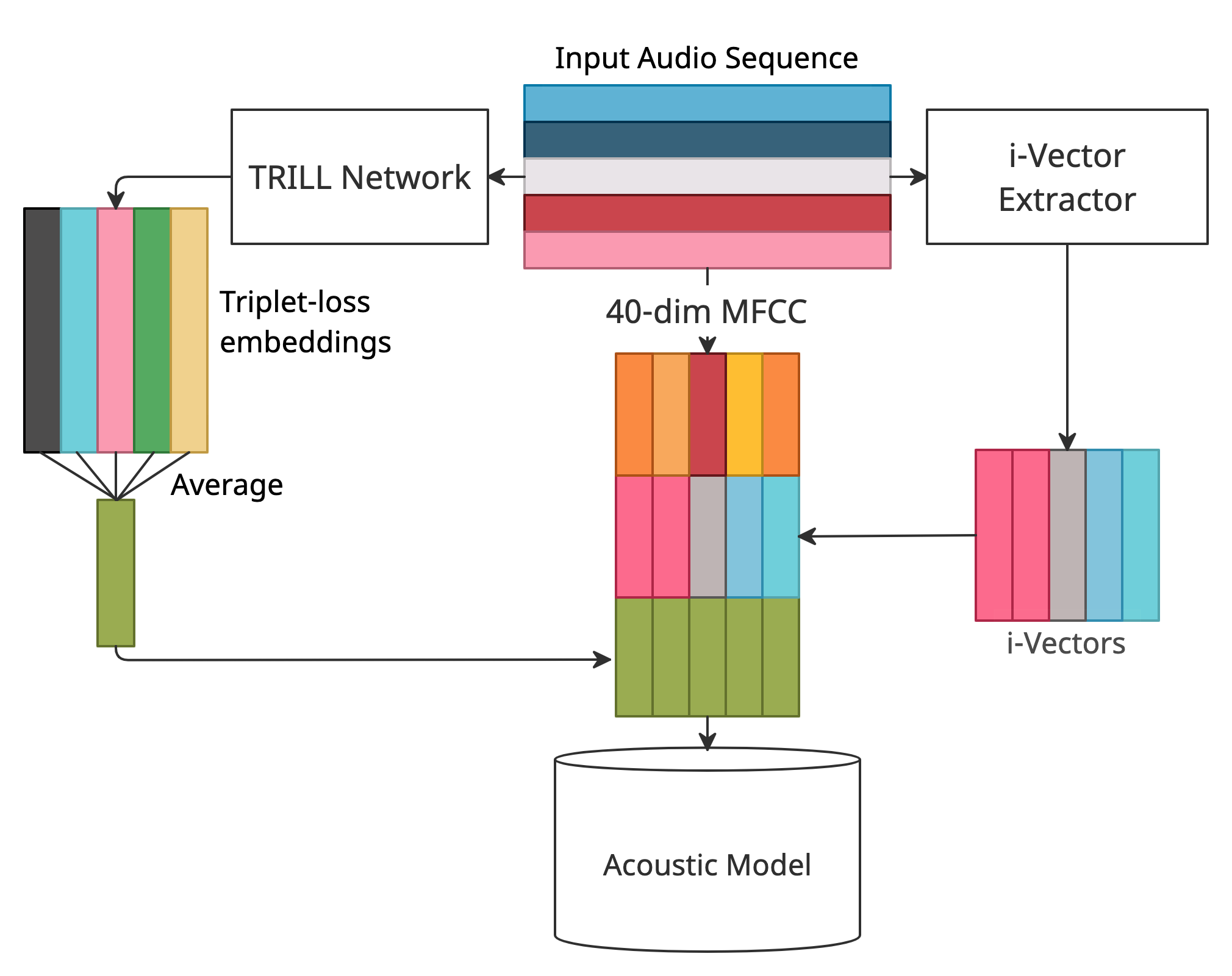}
  \caption{Feature generation flow chart}
  \label{fig:flow}
\end{figure}

The proposed method is assessed using two datasets, the 100-hour challenge corpus of the CRSS-UTDallas Fearless Steps Corpus (Sec.\ref{sec:FS_exp}) and CHiME-4 corpus (Sec.\ref{sec:c4_exp}). Systems development employs the Kaldi speech recognition toolkit \cite{povey2011kaldi}, and uses the same feature extraction pipeline shown in Fig.\ref{fig:flow}. For the CRSS-UTDallas Fearless Steps task, we utilize a factorized time delay neural network (TDNN-f) \cite{povey2018semi} for acoustic modeling while for CHiME-4, we focus on the 1 channel track task as they employed in \cite{chen2018building}.

\vspace{-0.12cm}
\section{Related work}
Historically, many approaches have been proposed to address noise robustness in ASR systems \cite{li2014overview, hansen2014environment}.
In \cite{seltzer2013investigation}, an approach based on noise-aware training which incorporates information about the environment was introduced into DNN training.
In \cite{qian2016neural}, three extraction models for speaker, phone, and environment were considered, along with a multi-task joint training architecture.
In \cite{liang2018learning}, the invariant representation learning technique was proposed, which demonstrated significant reduction in character error rate and robustness for out-of-domain noise settings.
In \cite{raj2020frustratingly}, a simple method was considered to extract a noise vector for acoustic model training. It is suggested that the technique could also be applied in online ASR by estimating the mean vector with frame-level maximum likelihood.
\vspace{-0.3cm}
\vspace{-0.6cm}

\section{Proposed system}
Given the challenges in robustness for ASR with CRSS-UTDallas Fearless Steps and CHiME-4, this section presents the formulation of our scenario aware based acoustic modeling to address environmental variability.

\subsection{Scenario Aware}
Factor aware training has been shown to be effective in ASR system development \cite{seltzer2013investigation, qian2016neural, raj2020frustratingly}. This training strategy produces a system that is more robust to factors such as noise, speaker, and room characteristics. Most earlier studies have used a representation for each specific distortion factor, where the extracted representations are either fed into the input layer, the hidden layer, or the output layer. In our study, we use a single representation to characterize all factors/acoustic info within an audio, including speech, which leads to a scenario aware training for the resulting acoustic model.

The input feature for our acoustic model contains two types of vectors, the first is the commonly used MFCC, along with the i-Vector which we denote as $m \in R ^{1 \times M}$. The second feature is the triplet-loss vector from TRILL, which we denote as $n \in R ^{1 \times N}$. This total input vector $V$ is represented as the concatenation of $M$ dimensional vector $m$ and $N$ dimensional vector $n$:
\begin{equation}
  V = [m;n] \in R ^ {1 \times (M + N)}
  \label{eq1}
\end{equation}
Note that we average the triplet-loss embeddings over time for an audio input to form a triplet-loss vector, that is one vector for the entire audio. A flow diagram is shown in Fig \ref{fig:flow}.

\vspace{-1em}
\subsection{Triplet-loss based Representation}
\label{sec:TRILL}
Previously, the triplet-loss based representation generated by TRILL model was introduced in \cite{shor2020towards} and originally used for non-semantic downstream tasks. The pre-trained model\footnote{https://tfhub.dev/google/nonsemantic-speech-benchmark/trill/3} we used was trained on a subset of AudioSet \cite{gemmeke2017audio} training set clips that own the speech label and with the ResNet-50 architecture discussed in \cite{hershey2017cnn}, followed by a 512-dimensional embedding layer.
Next, the temporal proximity is used as a self-supervision signal \cite{jansen2018unsupervised}. The idea behind using the temporal proximity is that sounds in a given environment are usually restricted to a subset of sound creating objects that are often closely related. Hence, a pair of events in the same audio should have a higher probability of being the same, or at least related in a certain level than any two audio clips randomly chosen from a large audio collection.

The collection of audio for training the triplet-loss model such as TRILL, can be depicted as a sequence of spectrogram context windows $X=[x_1,x_2,\dots,x_N]$, where $x_i \in R^{F \times T}$ with $F$ and $T$ represent frequency and time window. The goal for this model is to learn a map $g:R^{F \times T} \rightarrow R^D$ that transform $x_i$ into d-dimensional space such that $||g(x_i)-g(x_j)|| \leq ||g(x_i)-g(x_k)||$ when $|i-j| \leq |i-k|$. This is achieved by first sampling in $X$ a great amount of triplets $s=[x_i,x_j,x_k]$, which are known as the anchor, positive, and negative respectively, where $|i-j| \leq \tau$ and $|i-k| > \tau$ with a reasonable time scale $\tau$. Next, we train the model with the triplet loss:
\begin{equation}
  \mathcal{L}(s) = \sum_{i=1}^{N} [||g(x_i)-g(x_j)||^2_2 - ||g(x_i)-g(x_k)||^2_2 + \delta ]_+
  \label{eq2}
\end{equation}
where $||.||$ is the $L_2$ norm, $\delta$ is a non-negative margin hyperparameter, and $[.]_+$ is the hinge loss. It is clear the loss will be exactly zero if all the training triplets satisfy the inequality:
\begin{equation}
  ||g(x_i)-g(x_j)||^2_2 + \delta \leq ||g(x_i)-g(x_k)||^2_2
  \label{eq3}
\end{equation}

In \cite{shor2020towards}, the training task is based on a pair-wise data set with the same clip ($x_i,x_j$)/different clip ($x_i,x_k$) discrimination achieved by setting the utterance value $\tau$ to 10 seconds, which is the maximum duration of the clips in AudioSet. This makes the triplet-loss model capable of mapping audio clips with close acoustic information into similar embeddings. With the enormous size and scope of labels in AudioSet, we employed triplet-loss based representation to model the environment scenario in the audio.

\vspace{-0.2cm}
\section{Experiments}
The method for extracting the input features for the following experiments are all the same. Firstly, we extract 512 dimensional embeddings through the TRILL model and average these embeddings over time to form a triplet-loss vector. Next, the vector is combined with 40 dimensional MFCC features and 100 dimensional i-Vector as the input vector for acoustic model training. The feature generation pipeline is also shown in the Fig \ref{fig:flow}.

\vspace{-0.8em}
\subsection{Fearless Steps Corpus Experiments}
\label{sec:FS_exp}
\subsubsection{Data}
\label{sec:FS_data}
The Fearless Steps Corpus \cite{hansen2018fearless} consists of 19,000 hours of actual Apollo conversational speech across 30 time-synchronized channels, with Channel\#1 representing the time synchronized IRIG timecode. The audio represents actual communications from the Apollo-11 mission including all Mission Specialists, Astronauts, and support staff over the 7-day mission to the moon. These communication channel loops have distinct acoustic characteristics (e.g., noise, distortion, background interference, etc.) from NASA analog cables to the SoundScriber recording platform with channel and system noise that contributes to loss in ASR system performance. The variability across channel loops is due to the extensive cabling, headsets, relays, etc. necessary to bridge ~600 NASA specialists in different locations to allow them to communicate and work collaboratively to achieve a successful mission. All audio was recorded on 30-track analog 1 inch reel-to-reel recording tapes, and then digitized by CRSS-UTDallas initially at a 44.1kHz and later down-sampled to 8kHz, with 30 min. per data chunk for speech analysis.

For this study, we employ only the 100-hour Fearless Steps challenge corpus \cite{joglekar2020fearless} that consists of 5 selected channels with labeled data. This includes Network Controller (NTWK), Electrical, Environmental and Consumables Manager (EECOM), Guidance Navigation and Control (GNC), Flight Director (FD), Mission Operations Control Room (MOCR).
Here, we use the ASR track2 in the challenge corpus where the audio is already segmented with utterance level transcriptions. The training set is roughly 28 hours, with development set being 7.6 hours, and the evaluation set being 10.6 hours. The training set is used for both the acoustic and language model. We use the development set for computing the perplexity in language model training, and the evaluation set is used only for test.

\vspace{-0.7em}
\subsubsection{Baseline System}
\label{sec:FS_base}
For lexicon model, we employ the CMU dictionary\footnote{http://www.speech.cs.cmu.edu/cgi-bin/cmudict} as a basic pronunciation dictionary. However, since many words in the Fearless Steps Corpus are NASA space related and not present in the CMU dictionary, we use the Phonetisaurus G2P \cite{novak2012wfst} to generate pronunciations for these out-of-vocabulary words. A speaker adapted HMM-GMM is first trained on the training set to generate phoneme to audio alignments for DNN training. The TDNN-f \cite{povey2018semi} with 15 1024-dimensional layers factorized with 160-dimensional linear bottlenecks is used for acoustic modeling on the same dataset. For the language model, a basic 3-gram model was used, with pronunciation and silence probability modeling as described in \cite{chen2015pronunciation}.

\vspace{-0.7em}
\subsubsection{Results and Analysis}
\begin{figure}[th]
  \centering
  \includegraphics[width=7.8cm]{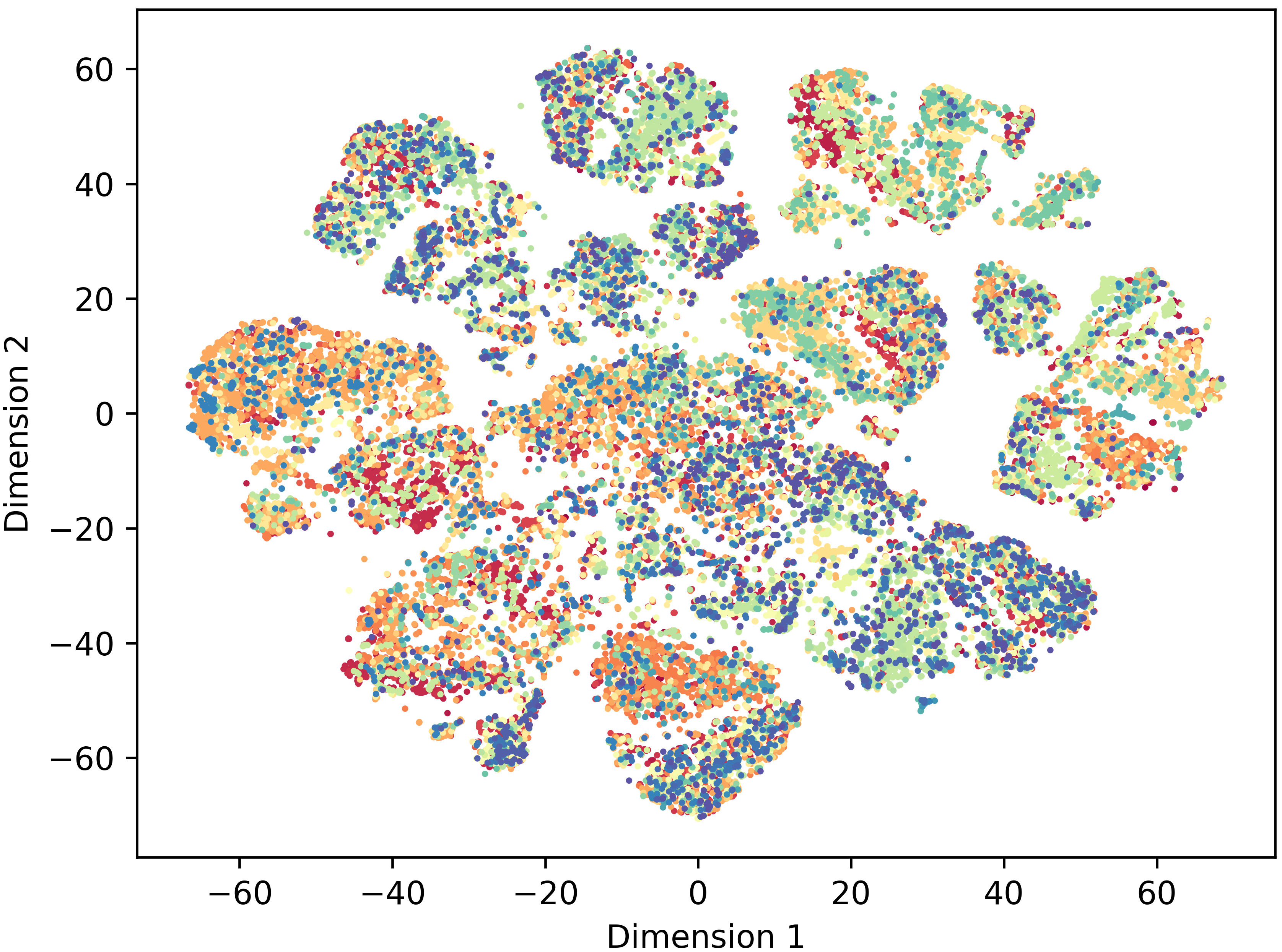}
  \vspace{-0.2cm}
  \caption{t-SNE of triplet-loss vectors based on TRILL model from Fearless Steps 100-hour challenge Corpus training set. The plot includes 35k utterances (points) with each color corresponding to one of 256 Apollo-11 speakers.}
  \label{fig:FS_TSNE}
\end{figure}

\begin{figure}[t]
  \centering
  \includegraphics[width=8cm]{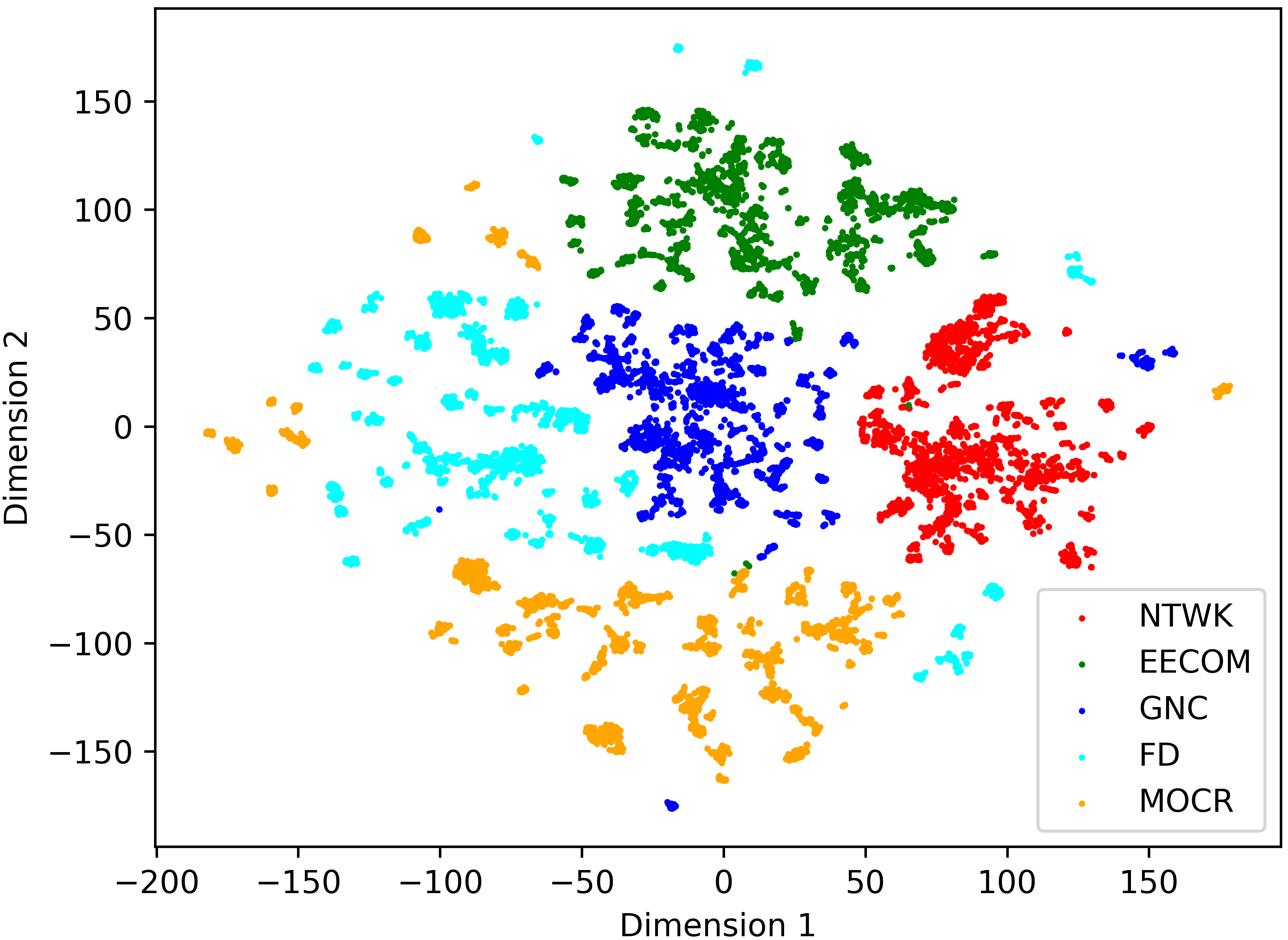}
  \vspace{-0.2cm}
  \caption{t-SNE of triplet-loss vectors based on TRILL model from a randomly selected 360-hour subset of Fearless Steps Apollo-11 corpus, with 5 of 30 possible channels analyzed with data being: NTWK:66hrs, EECOM:66hrs, GNC:66hrs, FD:81hrs, MOCR:81hrs.}
  \label{fig:A11_TSNE}
\end{figure}

The purpose of using a triplet-loss based representation, is to model the acoustic condition of each audio context. As shown in Fig.\ref{fig:FS_TSNE}, the extracted triplet-loss vectors from training data are categorized into a few different blocks. Each block can be treated as a distinct acoustic characteristic. Since we assume that the channel number corresponds to each utterance is not known, the speaker information is used instead as the label. Most speakers have utterances spread out across multiple blocks, but they do not necessarily cover every block.
The triplet-loss based representation allowed for the analysis of environment scenario as shown in Fig.\ref{fig:A11_TSNE}, which shows selected channels that have different characteristics. This figure used a randomly selected 360-hour subset of the complete 9,000-hour Fearless Steps Corpus that contains channel number information for each audio stream. Each point represents a triplet-loss vector extracted from TRILL model from a 15 second block of audio cut from the original 30 min. sequential audio chunks.

\begin{table}[h]
\caption{WER of Fearless Steps Corpus experiments. T-REP stands for triplet-loss based representation. \textit{Dim} is the layer dimension of TDNN-f.}
\label{tab:FS}
\centering
\scalebox{0.84}{
\begin{threeparttable}
  \begin{tabular}{l l c c c c}
    \toprule
    (No.) Model & i-vector & \textit{Dim} & \textbf{Dev (\%)} & \textbf{Eval (\%)}\\
   	\midrule
    (1) Baseline & yes & 1024 & 27.66 & 29.89 \\
    (2) Baseline\_1536 & yes & 1536 & 27.28 & 29.48 \\
    (3) + T-REP matrix* & yes & 1536 & 27.07 & 29.64\\
    (4) + T-REP & no & 1536 & 26.68 & 29.30\\
    (5) + T-REP & yes & 1536 & 26.49 & 29.17 \\
    (6) + T-REP \& multi-style & yes & 1536 & \textbf{26.16} & \textbf{28.94} \\
    \midrule
    Gorin et al.\cite{gorin400houston} & yes & 1024 & 28.60 & 31.4 \\
    \bottomrule
  \end{tabular}
  \begin{tablenotes}
    \item *This means we do not average over time on the embeddings.
  \end{tablenotes}
\end{threeparttable}}
\end{table}

In Table \ref{tab:FS}, word error rates (WER) are shown for experiments based on the Fearless Steps Corpus. The first row is the baseline system described in Sec.\ref{sec:FS_base}. We found that by increasing the layer dimension of the TDNN-f to 1536 (No.2), can further reduce WER, but increasing to larger dimensions such as 2136 (not shown here), caused a loss in performance. After adding the triplet-loss based representation, we observe a 2.9\% and 1.1\% relative WER improvement in development and evaluation set (No.5 vs. No.2). 
Another discovery is that using an average over time with triplet-loss embeddings advances the WER (No.5 vs. No.3). No.4 is the only experiment without the i-Vector. This shows that MFCC with the triplet-loss based representation (No.4) is better than MFCC with i-Vector (No.2). In all, our best system (No.6) with the triplet-loss based representation and multi-style training achieves 5.42\% and 3.18\% relative improvement on WER in development and evaluation set respectively. The multi-style training is accomplished by adding data augmented with the room impulse response (RIR) and MUSAN corpus (music, speech, and noise). With the original data included, we are able to expand the training set size by 5x vs. the original data.
Note that simply adding triplet-loss based representation provides more improvement than multi-style training (No.5 - No.2 vs. No.6 - No.5). We include the last row as a comparison to the best system in Fearless Steps Challenge Phase II, with the matched condition in both acoustic and language model and training data used in our system.

\vspace{-1em}
\subsection{CHiME-4 corpus Experiments}
\label{sec:c4_exp}
\subsubsection{Data}
The CHiME-4 data \cite{vincent2017analysis} includes real data recorded in real-world noisy environments, and simulated data, that is artificially created using clean speech data mixed with noisy background data. Five locations (i.e. booth (BTH), on the bus (BUS), cafe (CAF), pedestrian area (PED), and street junction (STR)) are chosen for real data recording. The BTH recordings are used for generating the simulated data, while all the rest are for ASR evaluation.

\vspace{-1em}
\subsubsection{Baseline System}
\label{sec:c4_base}
For a fair comparison, we only focus on the single channel track in the CHiME-4 challenge. The baseline system follows the work in \cite{chen2018building}, which uses a TDNN LF-MMI training on all 6 channels data and a LSTMLM trained with Kaldi-RNNLM \cite{xu2018neural} on a 3-fold texts of training data. The pronunciation dictionary was also based on the CMU dictionary.

\begin{table}[h]
  \caption{WER of CHiME-4 1 channel track experiments. T-REP stands for triplet-loss based representation.}
  \label{tab:1ch}
  \centering
  \begin{tabular}{ l c c c c }
    \toprule
    \multirow{2}{*}{\centering Model} & \multicolumn{2}{c}{\textbf{Dev (\%)}} & \multicolumn{2}{c}{\textbf{Test (\%)}} \\
    & real & simu & real & simu \\
   	\midrule
    Baseline & 5.28 & \textbf{6.52} & 11.68 & \textbf{11.98} \\
    + T-REP & \textbf{4.43} & 6.55 & \textbf{10.29} & 12.03 \\
    \bottomrule
  \end{tabular}
\end{table}

\begin{table*}[t]
  \caption{WER by location of the test set in CHiME-4 1 channel track experiments. T-REP stands for triplet-loss representation.}
  \label{tab:location}
  \centering
  \begin{tabular}{ l c c c c | c c c c}
    \toprule
    \multirow{2}{*}{\centering Model} & \multicolumn{4}{c}{\textbf{real (\%)}} & \multicolumn{4}{c}{\textbf{simu (\%)}} \\
    & BUS & CAFE & PED & STR & BUS & CAFE & PED & STR \\
   	\midrule
    Baseline &  18.81 & 11.88 & 8.46 & 7.56 & 8.72 & \textbf{13.80} & \textbf{11.49} & 13.90 \\
    + T-REP & \textbf{15.40} & \textbf{11.3} & \textbf{8.09} & \textbf{6.37} & \textbf{8.44} & 14.66 & 11.56 & \textbf{13.45} \\
    \bottomrule
    relative improvement & 18.13 & 4.88 & 4.37 & 15.74 & 3.21 & -6.23 & -0.61 & 3.24 \\
    \bottomrule
  \end{tabular}
\end{table*}
\vspace{-1em}
\begin{figure}[h]
  \centering
  \includegraphics[width=7.5cm]{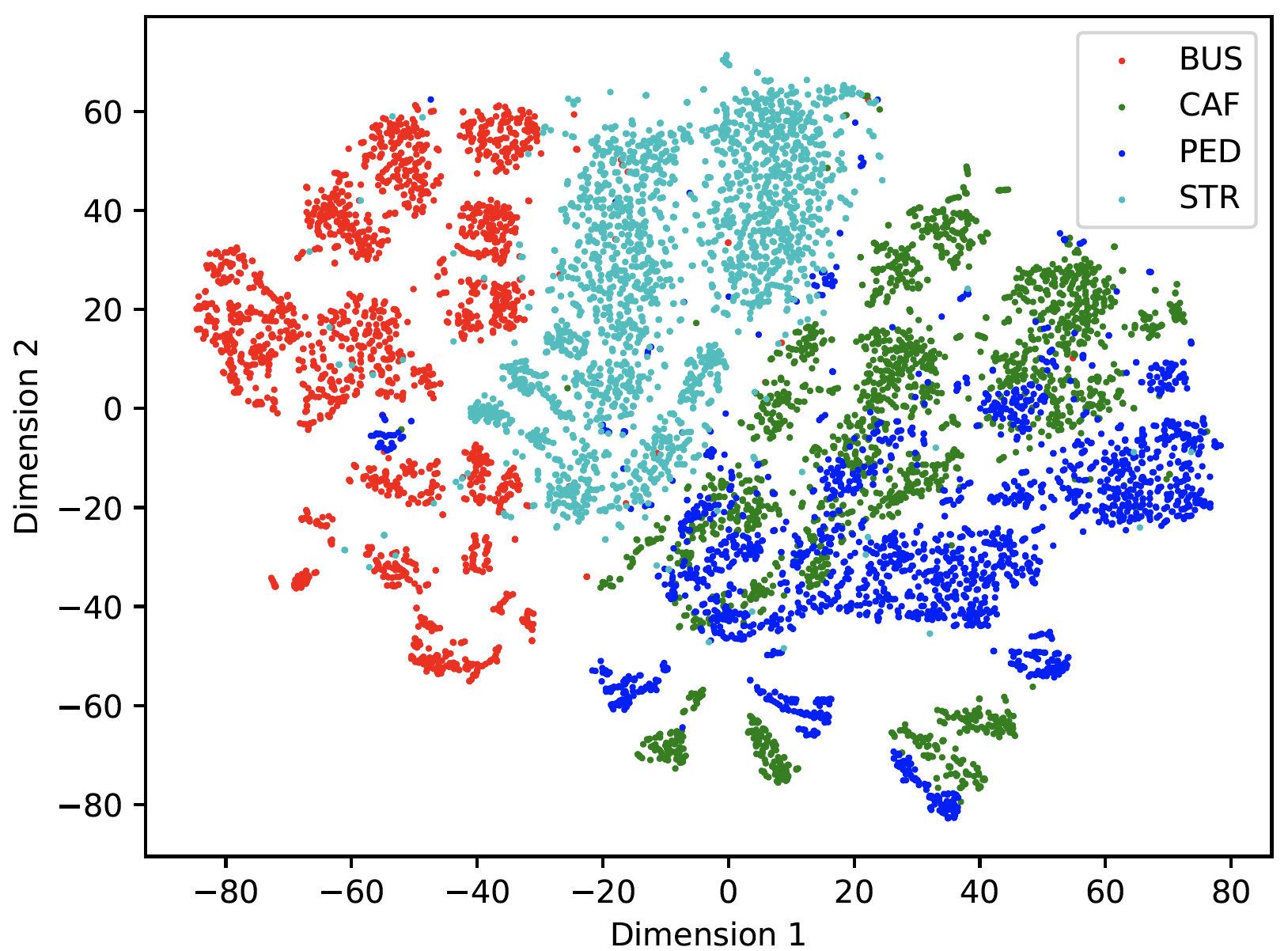}
  \vspace{-0.3cm}
  \caption{t-SNE of triplet-loss vectors based on TRILL model from CHiME-4 Corpus training set. The plot includes 8738 utterances (points).
  \vspace{-2ex}
  }
  \label{fig:chime4_TSNE}
\end{figure}

\vspace{-1em}
\subsubsection{Results and Analysis}
In Table \ref{tab:1ch}, we demonstrate the effect of adding the triplet-loss based representation. We observe a 16.10\% and 11.90\% relative WER improvement in real development and real test data by adding the triplet-loss based representation with only a small loss in simulation data.

The Fig.\ref{fig:chime4_TSNE} shows the t-SNE plot of triplet-loss vectors extracted from training set of CHiME-4 corpus. We can see a clear separation of each location, where only the CAF and PED are more overlapped than others. This observation matches the WER improvement. Also, it is suggested that this is one of the reasons why CHiME-4 benefits more from triplet-loss based representation versus the Fearless Steps Corpus.

We further investigate the effectiveness of triplet-loss based representation using the environment difference in the test set. In Table \ref{tab:location}, we show a greater improvement for real data versus simulation data. With the observation in Fig.\ref{fig:chime4_TSNE} in mind, it is shown that BUS and STR environment locations have greater improvement than CAFE and PED environment locations. This leads to the conclusion that the more distinct the acoustic context is in the audio from others, the more beneficial triplet-loss based representation will help.

\vspace{-1em}
\subsection{Analysis on Triplet-loss Representation Performance}
\label{sec:cross_analysis}
We note that there is a wide gap in performance of triplet-loss based representation between Fearless Steps and CHiME-4 corpora ( i.e. 1.1\% compared to 11.90\% relative improvements on WER). It is suggested that this is based on the dissimilar in formation of the acoustics for the difference in system improvements. As mentioned in Sec.\ref{sec:FS_data}, the audio for Fearless Steps are all analog recordings, where audio cable routing and channel recording conditions cause additional background noise that become a distinct characteristic of each channel. However, the specific distortion of these channels, such as strong low frequency harmonics, are highly unsophisticated, compared to the diverse general background noises in CHiME-4 data. Another possible reason is that the audio conversational-turn duration in Fearless Steps Corpus are commonly short, with a mean of 1.93 sec, and standard deviation of 3.28; while audio duration in CHiME-4 has a mean of 7.44 sec and standard deviation of 2.86. Here, 22\% of the training set consists of audio turns which are less than 1 sec. These short duration make it hard for triplet-loss based representation to be as meaningful or effective.

\vspace{-1em}
\section{Conclusion}
This study has considered a triplet-loss approach as our proposed method for scenario aware speech recognition. To employ triplet-loss based representation, we utilize the TRILL model (Sec.3.2) to model all factors/acoustic info within an utterance, leading to a scenario aware ASR system. This technique is especially beneficial for real data when compared to simulation data. Furthermore, the more distinct the background acoustic structure is from each other, the greater the improvement possible. The system achieved 5.42\% and 3.18\% relative WER improvement on the development and evaluation test sets of the Fearless Steps Corpus, and 11.90\% relative WER improvement on real test data of CHiME-4 corpus.

Our future work will explore alternative representations trained for different architectures and data. Also, we will further explore the integration between neural embeddings and the resulting acoustic model.
\vspace{-1em}
\section{Acknowledgements}
The authors would like to express our sincere thanks for valuable discussions with Wei-Cheng Lin, Midia Yousefi, and Aditya Joglekar. 


\bibliographystyle{IEEEbib}
\bibliography{refs}

\begin{thebibliography}{10}

\bibitem{bhattacharjee2016statistical}
Utpal Bhattacharjee, Swapnanil Gogoi, and Rubi Sharma,
\newblock ``A statistical analysis on the impact of noise on mfcc features for
  speech recognition,''
\newblock in {\em 2016 International Conference on Recent Advances and
  Innovations in Engineering (ICRAIE)}. IEEE, 2016, pp. 1--5.

\bibitem{gong1995speech}
Yifan Gong,
\newblock ``Speech recognition in noisy environments: A survey,''
\newblock {\em Speech communication}, vol. 16, no. 3, pp. 261--291, 1995.

\bibitem{yu2008minimum}
Dong Yu, Li~Deng, Jasha Droppo, Jian Wu, Yifan Gong, and Alex Acero,
\newblock ``A minimum-mean-square-error noise reduction algorithm on
  mel-frequency cepstra for robust speech recognition,''
\newblock in {\em 2008 IEEE International Conference on Acoustics, Speech and
  Signal Processing}. IEEE, 2008, pp. 4041--4044.

\bibitem{seltzer2010acoustic}
Michael~L Seltzer, Alex Acero, and Kaustubh Kalgaonkar,
\newblock ``Acoustic model adaptation via linear spline interpolation for
  robust speech recognition,''
\newblock in {\em 2010 IEEE International Conference on Acoustics, Speech and
  Signal Processing}. IEEE, 2010, pp. 4550--4553.

\bibitem{seltzer2013investigation}
Michael~L Seltzer, Dong Yu, and Yongqiang Wang,
\newblock ``An investigation of deep neural networks for noise robust speech
  recognition,''
\newblock in {\em 2013 IEEE international conference on acoustics, speech and
  signal processing}. IEEE, 2013, pp. 7398--7402.

\bibitem{tan2016speaker}
Tian Tan, Yanmin Qian, Dong Yu, Souvik Kundu, Liang Lu, Khe~Chai Sim, Xiong
  Xiao, and Yu~Zhang,
\newblock ``Speaker-aware training of lstm-rnns for acoustic modelling,''
\newblock in {\em 2016 IEEE International Conference on Acoustics, Speech and
  Signal Processing (ICASSP)}. IEEE, 2016, pp. 5280--5284.

\bibitem{giri2015improving}
Ritwik Giri, Michael~L Seltzer, Jasha Droppo, and Dong Yu,
\newblock ``Improving speech recognition in reverberation using a room-aware
  deep neural network and multi-task learning,''
\newblock in {\em 2015 IEEE International Conference on Acoustics, Speech and
  Signal Processing (ICASSP)}. IEEE, 2015, pp. 5014--5018.

\bibitem{dehak2010front}
Najim Dehak, Patrick~J Kenny, R{\'e}da Dehak, Pierre Dumouchel, and Pierre
  Ouellet,
\newblock ``Front-end factor analysis for speaker verification,''
\newblock {\em IEEE Transactions on Audio, Speech, and Language Processing},
  vol. 19, no. 4, pp. 788--798, 2010.

\bibitem{zhang2018text}
Chunlei Zhang, Kazuhito Koishida, and John~HL Hansen,
\newblock ``Text-independent speaker verification based on triplet
  convolutional neural network embeddings,''
\newblock {\em IEEE/ACM Transactions on Audio, Speech, and Language
  Processing}, vol. 26, no. 9, pp. 1633--1644, 2018.

\bibitem{zhang2019utd}
Chunlei Zhang, Fahimeh Bahmaninezhad, Shivesh Ranjan, Harishchandra Dubey, Wei
  Xia, and John~HL Hansen,
\newblock ``{UTD-CRSS} systems for 2018 {NIST} speaker recognition
  evaluation,''
\newblock in {\em ICASSP 2019-2019 IEEE International Conference on Acoustics,
  Speech and Signal Processing (ICASSP)}. IEEE, 2019, pp. 5776--5780.

\bibitem{shor2020towards}
Joel Shor, Aren Jansen, Ronnie Maor, Oran Lang, Omry Tuval, Felix de~Chaumont
  Quitry, Marco Tagliasacchi, Ira Shavitt, Dotan Emanuel, and Yinnon Haviv,
\newblock ``Towards learning a universal non-semantic representation of
  speech,''
\newblock {\em Proc. Interspeech 2020}, pp. 140--144, 2020.

\bibitem{gemmeke2017audio}
Jort~F Gemmeke, Daniel~PW Ellis, Dylan Freedman, Aren Jansen, Wade Lawrence,
  R~Channing Moore, Manoj Plakal, and Marvin Ritter,
\newblock ``Audio set: An ontology and human-labeled dataset for audio
  events,''
\newblock in {\em 2017 IEEE International Conference on Acoustics, Speech and
  Signal Processing (ICASSP)}. IEEE, 2017, pp. 776--780.

\bibitem{jansen2018unsupervised}
Aren Jansen, Manoj Plakal, Ratheet Pandya, Daniel~PW Ellis, Shawn Hershey,
  Jiayang Liu, R~Channing Moore, and Rif~A Saurous,
\newblock ``Unsupervised learning of semantic audio representations,''
\newblock in {\em 2018 IEEE international conference on acoustics, speech and
  signal processing (ICASSP)}. IEEE, 2018, pp. 126--130.

\bibitem{povey2011kaldi}
Daniel Povey, Arnab Ghoshal, Gilles Boulianne, Lukas Burget, Ondrej Glembek,
  Nagendra Goel, Mirko Hannemann, Petr Motlicek, Yanmin Qian, Petr Schwarz,
  et~al.,
\newblock ``The kaldi speech recognition toolkit,''
\newblock in {\em IEEE 2011 workshop on automatic speech recognition and
  understanding}. IEEE Signal Processing Society, 2011, number CONF.

\bibitem{povey2018semi}
Daniel Povey, Gaofeng Cheng, Yiming Wang, Ke~Li, Hainan Xu, Mahsa Yarmohammadi,
  and Sanjeev Khudanpur,
\newblock ``Semi-orthogonal low-rank matrix factorization for deep neural
  networks.,''
\newblock in {\em Interspeech}, 2018, pp. 3743--3747.

\bibitem{chen2018building}
Szu-Jui Chen, Aswin~Shanmugam Subramanian, Hainan Xu, and Shinji Watanabe,
\newblock ``Building state-of-the-art distant speech recognition using the
  {CHiME}-4 challenge with a setup of speech enhancement baseline,''
\newblock {\em arXiv preprint arXiv:1803.10109}, 2018.

\bibitem{li2014overview}
Jinyu Li, Li~Deng, Yifan Gong, and Reinhold Haeb-Umbach,
\newblock ``An overview of noise-robust automatic speech recognition,''
\newblock {\em IEEE/ACM Transactions on Audio, Speech, and Language
  Processing}, vol. 22, no. 4, pp. 745--777, 2014.

\bibitem{hansen2014environment}
John~HL Hansen, Abhishek Kumar, and Pongtep Angkititrakul,
\newblock ``Environment mismatch compensation using average eigenspace-based
  methods for robust speech recognition,''
\newblock {\em International Journal of Speech Technology}, vol. 17, no. 4, pp.
  353--364, 2014.

\bibitem{qian2016neural}
Yanmin Qian, Tian Tan, and Dong Yu,
\newblock ``Neural network based multi-factor aware joint training for robust
  speech recognition,''
\newblock {\em IEEE/ACM Transactions on Audio, Speech, and Language
  Processing}, vol. 24, no. 12, pp. 2231--2240, 2016.

\bibitem{liang2018learning}
Davis Liang, Zhiheng Huang, and Zachary~C Lipton,
\newblock ``Learning noise-invariant representations for robust speech
  recognition,''
\newblock in {\em 2018 IEEE Spoken Language Technology Workshop (SLT)}. IEEE,
  2018, pp. 56--63.

\bibitem{raj2020frustratingly}
Desh Raj, Jesus Villalba, Daniel Povey, and Sanjeev Khudanpur,
\newblock ``Frustratingly easy noise-aware training of acoustic models,''
\newblock {\em arXiv preprint arXiv:2011.02090}, 2020.

\bibitem{hershey2017cnn}
Shawn Hershey, Sourish Chaudhuri, Daniel~PW Ellis, Jort~F Gemmeke, Aren Jansen,
  R~Channing Moore, Manoj Plakal, Devin Platt, Rif~A Saurous, Bryan Seybold,
  et~al.,
\newblock ``{CNN} architectures for large-scale audio classification,''
\newblock in {\em 2017 ieee international conference on acoustics, speech and
  signal processing (icassp)}. IEEE, 2017, pp. 131--135.

\bibitem{hansen2018fearless}
John~HL Hansen, Abhijeet Sangwan, Aditya Joglekar, Ahmet~Emin Bulut, Lakshmish
  Kaushik, and Chengzhu Yu,
\newblock ``Fearless steps: Apollo-11 corpus advancements for speech
  technologies from earth to the moon.,''
\newblock in {\em INTERSPEECH}, 2018, pp. 2758--2762.

\bibitem{joglekar2020fearless}
Aditya Joglekar, John~HL Hansen, Meena~Chandra Shekar, and Abhijeet Sangwan,
\newblock ``Fearless steps challenge ({FS}-2): Supervised learning with massive
  naturalistic apollo data,''
\newblock {\em Proc. Interspeech 2020}, pp. 2617--2621, 2020.

\bibitem{novak2012wfst}
Josef~R Novak, Nobuaki Minematsu, and Keikichi Hirose,
\newblock ``{WFST}-based grapheme-to-phoneme conversion: Open source tools for
  alignment, model-building and decoding,''
\newblock in {\em Proceedings of the 10th International Workshop on Finite
  State Methods and Natural Language Processing}, 2012, pp. 45--49.

\bibitem{chen2015pronunciation}
Guoguo Chen, Hainan Xu, Minhua Wu, Daniel Povey, and Sanjeev Khudanpur,
\newblock ``Pronunciation and silence probability modeling for asr,''
\newblock in {\em Sixteenth Annual Conference of the International Speech
  Communication Association}, 2015.

\bibitem{gorin400houston}
Arseniy Gorin, Daniil Kulko, Steven Grima, and Alex Glasman,
\newblock ``{“This is Houston. Say again, please”}. the behavox system for
  the {Apollo-11 Fearless Steps Challenge (Phase II).},''
\newblock {\em Power}, vol. 400, no. 800, pp. 1200.

\bibitem{vincent2017analysis}
Emmanuel Vincent, Shinji Watanabe, Aditya~Arie Nugraha, Jon Barker, and Ricard
  Marxer,
\newblock ``An analysis of environment, microphone and data simulation
  mismatches in robust speech recognition,''
\newblock {\em Computer Speech \& Language}, vol. 46, pp. 535--557, 2017.

\bibitem{xu2018neural}
Hainan Xu, Ke~Li, Yiming Wang, Jian Wang, Shiyin Kang, Xie Chen, Daniel Povey,
  and Sanjeev Khudanpur,
\newblock ``Neural network language modeling with letter-based features and
  importance sampling,''
\newblock in {\em 2018 IEEE international conference on acoustics, speech and
  signal processing (ICASSP)}. IEEE, 2018, pp. 6109--6113.

\end{thebibliography}

\end{document}